\documentclass[11pt,showpacs,preprint,preprintnumbers,nofootinbib,superscriptaddress,amsmath,amssymb]{revtex4}

\def\be{\begin{equation}}
\def\ee{\end{equation}}
\newcommand{\bea}{\begin{eqnarray}}
\newcommand{\eea}{\end{eqnarray}}
\newcommand{\nn}{\nonumber}

\usepackage{graphicx}
\usepackage{dcolumn}
\usepackage{bm}
\usepackage{amssymb}
\usepackage{amsmath}
\usepackage{epsfig}    
\usepackage{color}
\usepackage{slashed}
\newcolumntype{I}{!{\vrule width 1.3pt}}
\begin{document} 
\title{Radiative Generation of Lepton Masses with the $U(1)'$ Gauge Symmetry}
\preprint{KIAS-P14032}

\author{Hiroshi Okada}
\email{hokada@kias.re.kr}
\affiliation{School of Physics, KIAS, Seoul 130-722, Korea}
\author{Kei Yagyu}
\email{keiyagyu@ncu.edu.tw}
\affiliation{Department of Physics and Center for Mathematics and Theoretical Physics,
National Central University, Chungli, Taiwan 32001, ROC}

\begin{abstract}

We revisit our previous model proposed in Ref.~\cite{Okada:2013iba}, in which 
lepton masses except the tauon mass are generated at the one-loop level in a TeV scale physics. 
Although in the previous work, 
rather large Yukawa couplings constants; i.e., greater than about 3, 
are required to reproduce the muon mass, 
we do not need to introduce such a large but ${\cal O}$(1) couplings. 
In our model, masses for neutrinos (charged-leptons) are generated by a dimension five effective operator with 
two isospin triplet (singlet and doublet) scalar fields. 
Thus, the mass hierarchy between neutrinos and charged-leptons can be naturally described 
by the difference in the number of vacuum expectation values (VEVs) 
of the triplet fields which must be much smaller than the VEV of the doublet field due to the constraint from the electroweak rho parameter. 
Furthermore, the discrepancy in the measured muon anomalous magnetic moment ($g-2$) from the prediction in the standard model are 
explained by one-loop contributions from vector-like extra charged-leptons which are necessary to obtain the radiative generation of the lepton masses. 
We study the decay property of the extra leptons by taking into account the masses of muon, neutrinos, muon $g-2$ and 
dark matter physics. We find that the extra leptons can mainly decay into the mono-muon, dark matter with or without $Z$ bosons in the favored parameter regions. 

\end{abstract}
\maketitle

\section{Introduction}

The standard model (SM) can successfully describe almost all the phenomena at collider experiments even after
the discovery of the Higgs boson at the LHC~\cite{Higgs}. 
However, it is well known that there are phenomena which cannot be explained in the SM such as the neutrino oscillations, 
the existence of dark matter (DM) and baryon asymmetry of the Universe. 
This strongly suggests that the SM should be replaced by a new physics model giving an explanation of these phenomena. 

One of the attractive scenarios to explain tiny neutrino masses 
is obtained in radiative seesaw models, in which the dimension five operator $\overline{L_L^c}L_L\Phi\Phi$, where
$L_L$ and $\Phi$ are respectively the left-handed lepton doublet and the Higgs doublet fields, 
supplying Majorana type neutrino masses is generated through quantum levels. 
Thanks to a loop suppression factor, a new physics scale typically described by masses of new particles running in the loop 
can be of order 1 TeV. 
Therefore, direct searches for this class of models are possible at collider experiments. 
Furthermore, a DM candidate can be naturally obtained\footnote{There are other types of radiative seesaw models without containing a DM candidate; e.g., 
the Zee model~\cite{Zee} and the Zee-Babu model~\cite{Zee-Babu}. } 
due to an unbroken discrete symmetry which is necessary to enclose a loop diagram generating neutrino masses and to forbid lower order masses such as a tree level Dirac neutrino mass.

So far, various models have been constructed in this line. 
The model by Krauss, Nasri and Trodden has been proposed in the very early stage~\cite{KNT_model,KNT_model_ext}, 
in which neutrino masses are generated at the three-loop level, and its phenomenology at $e^+e^-$ colliders has been discussed in Refs.~\cite{KNT_model_pheno}. 
Another simple model with one-loop induced neutrino masses has been constructed by Ma~\cite{Ma_model,Ma_model_DM,Ma_model_RGE}, 
and its extensions have also been discussed in Refs.~\cite{Ma_model_ext}. 
The model by Aoki, Kanemura and Seto~\cite{AKS_model,AKS2,AKS3} is the three-loop radiative seesaw model, where
the strong first order electroweak phase transition and additional CP phases in the Higgs sector 
can be realized, which is required by the successful electroweak baryogenesis scenario~\cite{EWBG}. 
Models with radiative generations for Dirac type masses for neutrinos have been proposed in Refs.~\cite{Radseesaw_Dirac}. 
In addition to the above modes, there are a lot of papers proposing various types of radiative seesaw model~\cite{Radseesaw_variation, Radseesaw_Non-Abelian}. 
	
Apart from neutrino masses, 
the masses of charged-leptons are also so small compared to the electroweak scale; i.e., order of 100 GeV, especially 
the muon and electron masses. 
In the SM, 
smallness of the charged-lepton masses is just accommodated by taking the Yukawa coupling constants to be $\mathcal{O}(10^{-3})$ 
and $\mathcal{O}(10^{-5})$ for the muon and electron masses, respectively. 
In Refs.~\cite{1loop_fermion, Penedo}, several models have been proposed, where charged-lepton masses are radiatively induced\footnote{In Ref.~\cite{Ibarra}, the quark masses and mixings are radiatively induced in a model with two Higgs doublet Higgs fields. }.  
However, tiny neutrino masses are not explained simultaneously in a given model. 

In this paper, we would like to explain the following two questions regarding the lepton masses by extending 
the radiative seesaw mechanism; (1) why they are so small compared to 
the electroweak scale, and (2) why there is a large difference between masses of neutrinos and those of the electron or muon. 
In Ref.~\cite{Okada:2013iba}, we have proposed a new mechanism 
where Majorana masses of neutrinos and Dirac masses of charged-leptons are 
induced from the different type of dimension five operators; $\overline{L_L^c}L_L\Delta_0\Delta_1$ and $\bar{L}_Le_R^{}\Phi\Delta_0$, respectively, 
where $\Delta_0$ ($\Delta_1$) 
is a hypercharge\footnote{
The definition of the hypercharge $Y$ is given as $Q=Y+T_3$ with $Q$ and $T_3$ being the electromagnetic charge and the third component of the isospin. } 
$Y=0$ ($Y=1$) isospin triplet scalar field, and $e_R^{}$ 
is the right-handed charged-lepton singlet fields.  
It is known that the magnitude of the vacuum expectation value (VEV) of triplet scalar fields are severely constrained by
the electroweak rho parameter; i.e., they have to be smaller than order or 1 GeV. 
Therefore, for the equation (1), smallness can be explained by the loop suppression factor if the dimension five operators are 
generated via loop levels and the tiny VEVs of triplet scalar fields as well.  
In addition, the question (2) can be described by the difference in the number of triplet VEVs for the generation of masse for neutrinos and that for 
charged-leptons.   
 
We then have constructed a concrete renormalizable model~\cite{Okada:2013iba} incorporated the above mechanism, in which 
both the dimension five operators are induced at the one-loop level. 
However, we need rather large Yukawa coupling constants such as greater than about 3 to reproduce the muon mass. 
The main reason of this problem comes from the too strong suppression by the triplet VEV for the muon mass. 
Therefore, in this paper, 
we replace the one-loop induced operator $\bar{L}_Le_R^{}\Phi\Delta_0$ by $\bar{L}_Le_R^{}\Phi\chi$ with a SM gauge singlet scalar field $\chi$. 
We introduce an additional local $U(1)'$ symmetry which is spontaneously broken by the singlet VEV, so that 
the singlet VEV is expected to be order of 1 TeV with an order one $U(1)'$ gauge coupling constant to get a mass of extra gauge boson 
to be $\mathcal{O}$(1) TeV. 
Under these modifications, we can reproduce the muon mass with $\mathcal{O}$(1) Yukawa coupling constants. 
 
In our model, additional vector-like charged-leptons play a crucial role in generating the lepton masses. 
Moreover, the discrepancy in the observed muon anomalous magnetic moment (muon $g-2$) from the prediction in the SM can be compensated by the 
one-loop contributions of the additional charged-leptons with the mass of order 1 TeV. 
We discuss the decay property of the additional charged-leptons in the favored parameter regions 
by taking into account the masses of the muon, neutrinos, muon $g-2$ and dark matter physics.

This paper is organized as follows. 
In Sec.~II, we define our model, and we give the Lagrangian relevant to the generation of the lepton masses.
In Sec.~III, several observables in the lepton sector are calculated, e.g., masses for the charged-leptons and neutrinos, the
muon $g-2$, and lepton flavor violating (LFV) processes. 
Sec.~IV is devoted to study the decay property of the extra charged-leptons in the favored parameter regions. 
Conclusions and discussions are given in Sec.~V. Explicit formulae for the mass matrices for Higgs bosons are given in Appendix.

\section{The Model}

\begin{center}
\begin{table}[t]
\begin{tabular}{c|c|c|c|c|c}
\hline\hline Fermions & $L_L^i=(L_L^e,L_L^\mu,L_L^\tau)$ & $ e_R^a=(e_R^{},\mu_R^{}) $ & $ \tau_R^{} $  &  $E_L^\alpha$ & $E_R^\alpha$  \\\hline
$SU(2)_I,~U(1)_Y$ & $\bm{2},-1/2$ & $\bm{1},-1$& $\bm{1},-1$ & $\bm{1},-1$ & $\bm{1},-1$ \\\hline
$U(1)'$ & $x$ & $y$ & $x$ & $-x+2y$ & $-x+2y$   \\\hline
$\mathbb{Z}_2$ & $+$ & $+$ & $+$ & $-$ & $-$  \\\hline\hline
\end{tabular}\\ \vspace{0.6cm}
\begin{tabular}{c|c|c|c|c|c|c|c}
\hline\hline 
Scalar bosons &  $\Phi$ & $\Delta_0$  & $\Delta_1$  & $\eta$  & $\Phi_{3/2}$  & $S$ & $\chi$ \\\hline
$SU(2)_I,~U(1)_Y$ & $\bm{2},1/2$ & $\bm{3},0$  & $\bm{3},1$ & $\bm{2},1/2$  & $\bm{2},3/2$   & $\bm{1},0$  & $\bm{1},0$ \\\hline
$U(1)'$ & $0$ & $z$  & $-(2x+z)$ & $2(x-y)$ & $-2y$ & $y-x$  & $x-y$  \\\hline
$\mathbb{Z}_2$ & $+$ & $+$  & $+$ & $-$ & $-$ & $-$ & $+$ \\\hline\hline
\end{tabular}
\caption{The contents of lepton (upper table) and scalar boson (lower table) fields
and their charge assignment under $SU(2)_I\times U(1)_Y\times U(1)'\times\mathbb{Z}_2$, where 
$U(1)'$ is the additional gauge symmetry. 
The $U(1)'$ charges for $L_L^i$, $e_R^a$ and $\Delta_0$ are respectively denoted as $x$, $y$ and $z$, and 
those for all the other fields are expressed in terms of $x$, $y$ and $z$. 
The index $i$ ($a$) for $L_L$ ($e_R$) runs over the first, second and third (first and second) generation. }
\label{tab:1}
\end{table}
\end{center}

We propose a radiative lepton mass model where both Dirac charged-lepton (muon and electron) masses 
and Majorana neutrino masses are generated at the 1-loop level. 
We introduce an extra local $U(1)'$ (spontaneously broken) and a discrete $\mathbb{Z}_2$ (unbroken) 
symmetries in addition to the SM gauge symmetry. 
The particle contents and charge assignment 
are shown in TABLE~\ref{tab:1}. 
To avoid a tree level large mixing between the Z boson and a new $U(1)'$ gauge boson, we take the $U(1)'$ charge 
for the doublet Higgs field $\Phi$ to be zero\footnote{Although in general, there is a mixing from the gauge kinetic term, we just drop such a mixing term by hand.}.
Under the requirement where all the terms given in Eq.~(\ref{Yuk}) are allowed, 
all the other $U(1)'$ charges for the fields listed in TABLE~\ref{tab:1} can be written in terms of those for
$L_L^i$, $e_R^a$ and $\Delta_0$ denoted by $x$, $y$ and $z$, respectively.  
In order to forbid undesired terms giving tree level masses for the charged-leptons and neutrinos; i.e., 
$\overline{L^{i c}_L}\Delta_1 L_L^j$ and $\overline{L_L^i} \Phi e_R^{a}$, 
$z \neq0$ and $x-y\neq0$ must be satisfied, respectively. 
From the former condition, $\Delta_0$ has to be a complex field. 
Such a complex $Y=0$ triplet scalar field has also been introduced in Ref.~\cite{micchan} in a supersymmetric model. 
Notice here that the condition $x\neq y$ suggests that the $U(1)'$ symmetry cannot be identified as a lepton number symmetry.
The scalar fields $\Phi_{3/2}$, $\eta$ and $S$, and the vector-like charged-leptons $E^\alpha$
are assigned to be $\mathbb{Z}_2$-odd to enclose the loop in diagrams for the radiative generation of lepton masses. 

Comparing the current model with the previous our model, 
the SM gauge singlet scalar field $\chi$ with a non-zero VEV is additionally introduced, and the $Y=0$ triplet scalar field $\Delta_0$ is extended to be the complex field as mentioned in the above\footnote{We can also construct another model without the $\Delta_0$ field by changing the $U(1)'$ charge assignment 
for fields, in which the dimension five operator $\overline{L_L^c}L_L\Delta_0\Delta_1$ for neutrino masses is replaced by $\overline{L_L^c}L_L\Delta_0\Delta_1$}. 

\begin{figure}[t]
\begin{center}
 \includegraphics[width=65mm]{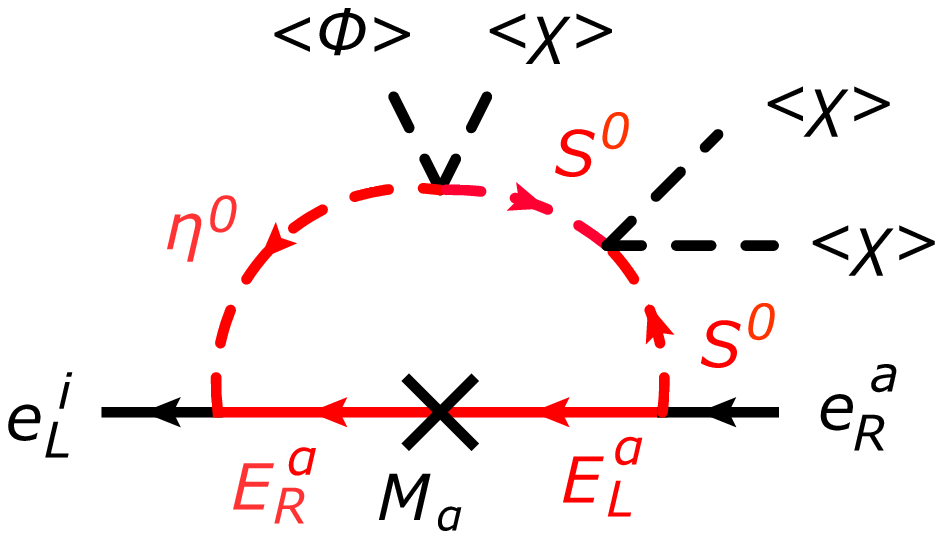}\\ \vspace{5mm}
 \includegraphics[width=130mm]{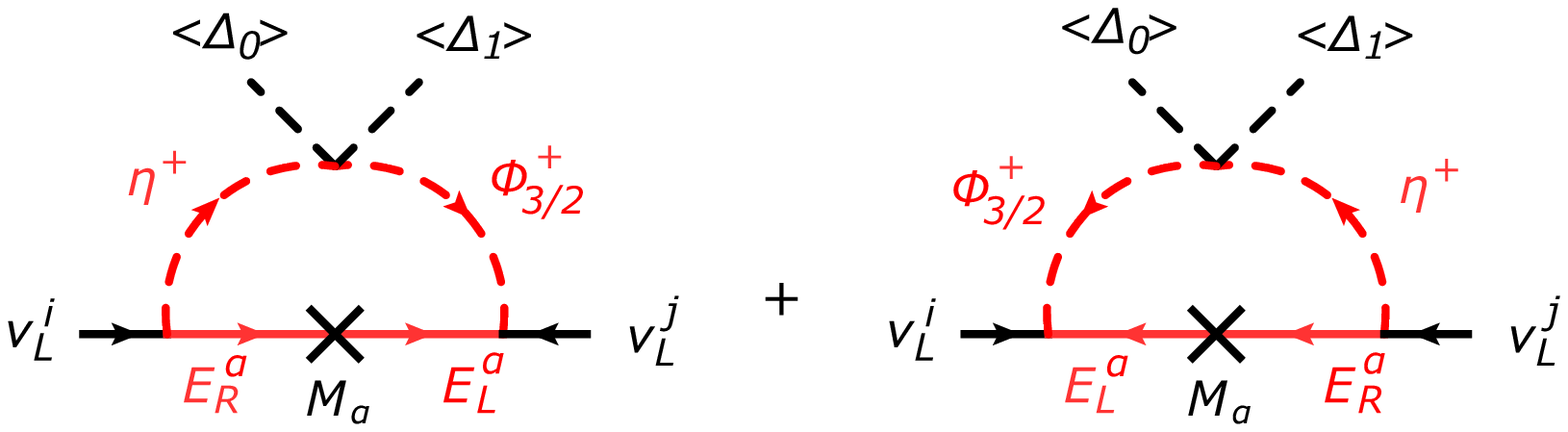}
   \caption{
Feynman diagrams for the one-loop generation of the charged-lepton masses (upper panel) and neutrino masses (lower panel). The particles indicated by the red font have the $\mathbb{Z}_2$-odd parity.}
   \label{Fig:Rad_Lep}
\end{center}
\end{figure}

The relevant Lagrangian to the radiative generations of lepton masses is given as follows
\begin{align}
-\mathcal{L}&= M_{\alpha}\overline{E_R^\alpha} E_L^\alpha+y_\tau^i\overline{L_L^i}\Phi\tau_R^{} +\rm{h.c.}\notag\\
&+y_S^{a\alpha}\overline{e_R^a}  E_L^\alpha S^*
+y_{\eta}^{i\alpha}\overline{L_L^i}\eta E_R^\alpha 
+y_{3/2}^{i\alpha} \overline{L_L^{ic}}(i\tau_2)\Phi_{3/2}E_L^\alpha +\rm{h.c.}\notag\\
&+\kappa_{e1} \chi^2S^2
+\kappa_{e2} \eta^\dagger  \Phi S^*\chi 
+\kappa_{\nu}  {\rm Tr}(\Delta_1\cdot\Delta_0) (\Phi_{3/2}^\dagger\cdot\eta)
+\text{h.c.}, 
 \label{Yuk}
\end{align}
where $M_{\alpha}$ is the mass of the $\alpha$-th vector-like lepton, and a pair of $\cdot$ appeared in the $\kappa_\nu$ term denotes the contraction by the Pauli matrices; i.e., $(A\cdot B)(C\cdot D)\equiv \sum_{i=1,2,3}(A\tau^i B)(C\tau^i D)$. 
In Fig.~\ref{Fig:Rad_Lep}, Feynman diagrams for the Dirac charged-lepton masses and Majorana neutrino masses are shown. 
Calculations of these diagrams are performed in the next section. 

The scalar potential can be separated into the $\mathbb{Z}_2$-even, $\mathbb{Z}_2$-odd and interaction parts due to the unbroken $\mathbb{Z}_2$ parity as
\begin{align}
V=V_{Z_2\text{-even}}+V_{Z_2\text{-odd}}+V_{\text{int}}, 
\end{align}
where each part is given by 
\begin{align}
V_{Z_2\text{-even}}&=
+m_\Phi^2 \Phi^\dagger \Phi
+m_{\Delta_1}^2\text{Tr}(\Delta_1^\dagger \Delta_1)
+m_{\Delta_0}^2\text{Tr}(\Delta_0^\dagger \Delta_0)
+m_\chi^2 \chi^*\chi\notag\\
& +\lambda_1(\Phi^\dagger \Phi)^2
+\lambda_2\left[\text{Tr}(\Delta_1^\dagger \Delta_1)\right]^2
+\lambda_3\text{Tr}(\Delta_1^\dagger \Delta_1)^2
+\lambda_4\left[\text{Tr}(\Delta_0^\dagger \Delta_0)\right]^2
+\lambda_5\text{Tr}(\Delta_0^\dagger \Delta_0)^2
+\lambda_6 (\chi^*\chi)^2\notag\\
& +\lambda_7(\Phi^\dagger \Phi)\text{Tr}(\Delta_1^\dagger \Delta_1)
+\lambda_8(\Phi^\dagger \cdot\Phi)\text{Tr}(\Delta_1^\dagger \cdot\Delta_1)
+\lambda_9(\Phi^\dagger \Phi)\text{Tr}(\Delta_0^\dagger \Delta_0)
+\lambda_{10}(\Phi^\dagger \cdot\Phi)\text{Tr}(\Delta_0^\dagger \cdot\Delta_0)\notag\\
&+\lambda_{11}(\Phi^\dagger \Phi)\chi^*\chi 
+\lambda_{12}\text{Tr}(\Delta_1^\dagger \Delta_1)\chi^*\chi 
+\lambda_{13} \text{Tr}(\Delta_0^\dagger \Delta_0)\chi^*\chi \notag\\
&+\lambda_{14}\text{Tr}(\Delta_1^\dagger \Delta_1)\text{Tr}(\Delta_0^\dagger \Delta_0)
+\lambda_{15}\text{Tr}(\Delta_1^\dagger \Delta_0)\text{Tr}(\Delta_0^\dagger \Delta_1)
+\lambda_{16}\text{Tr}(\Delta_1^\dagger \Delta_1\Delta_0^\dagger \Delta_0)\notag\\
&+\lambda_0\Phi^T(i\tau_2)\Delta_1^\dagger \Phi \chi
+\lambda'_0\Phi^\dagger\Delta_0 \Phi \chi+\text{h.c.},\label{Eq:Pot_Even}\\
V_{Z_2\text{-odd}}&=+m_\eta^2\eta^\dagger\eta +m_{3/2}^2\Phi_{3/2}^\dagger \Phi_{3/2}+m_{S}^2S^*S\notag\\
&+\xi_1 (\eta^\dagger \eta)^2+\xi_2 (\Phi_{3/2}^\dagger \Phi_{3/2})^2
+\xi_3 (S^*S )^2\notag\\
&+\xi_4(\eta^\dagger \eta)(\Phi_{3/2}^\dagger \Phi_{3/2})
+\xi_5|\eta^\dagger \Phi_{3/2}|^2
+\xi_6(\eta^\dagger \eta)S^*S, \label{Eq:Pot_Odd}\\
V_{\text{int}}&= +\kappa_1(\Phi^\dagger \Phi)(\eta^\dagger\eta)
+\kappa_2|\Phi^\dagger \eta|^2
+\kappa_3(\Phi^\dagger \Phi)(\Phi_{3/2}^\dagger \Phi_{3/2})
+\kappa_4|\Phi^\dagger \Phi_{3/2}|^2
+\kappa_5(\Phi^\dagger \Phi)S^*S\notag\\
&+\kappa_6 \text{Tr}(\Delta_1^\dagger\Delta_1)\eta^\dagger \eta
+\kappa_7 \text{Tr}(\Delta_1^\dagger\cdot \Delta_1)\eta^\dagger \cdot\eta
+\kappa_8 \text{Tr}(\Delta_1^\dagger\Delta_1)\Phi_{3/2}^\dagger \Phi_{3/2}
+\kappa_{9} \text{Tr}(\Delta_1^\dagger\cdot \Delta_1)\Phi_{3/2}^\dagger \cdot\Phi_{3/2}\notag\\
&+\kappa_{10} \text{Tr}(\Delta_1^\dagger\Delta_1)S^*S
+\kappa_{11} \text{Tr}(\Delta_0^\dagger\Delta_0)\eta^\dagger \eta
+\kappa_{12} \text{Tr}(\Delta_0^\dagger\cdot \Delta_0)\eta^\dagger \cdot \eta\notag\\
&
+\kappa_{13} \text{Tr}(\Delta_0^\dagger\Delta_0)\Phi_{3/2}^\dagger \Phi_{3/2}
+\kappa_{14} \text{Tr}(\Delta_0^\dagger\cdot \Delta_0)\Phi_{3/2}^\dagger \cdot \Phi_{3/2}\notag\\
&+\kappa_{15} \text{Tr}(\Delta_0^\dagger\Delta_0)S^*S
+\kappa_{16}\chi^*\chi \eta^\dagger \eta 
+\kappa_{17}\chi^*\chi \Phi_{3/2}^\dagger \Phi_{3/2}
+\kappa_{18}\chi^*\chi S^*S\notag\\
&
+\kappa_{e1} \chi^2S^2
+\kappa_{e2} \eta^\dagger  \Phi S^*\chi 
+\kappa_{\nu}  {\rm Tr}(\Delta_1\cdot\Delta_0) (\Phi_{3/2}^\dagger\cdot\eta)
+\tilde{\kappa}_{\nu}  {\rm Tr}(\Delta_1\Delta_0) (\Phi_{3/2}^\dagger\eta)
+\text{h.c.}.  \label{Eq:Pot_Int}
\end{align}
The $\kappa_{e1}$, $\kappa_{e2}$ and $\kappa_\nu$ terms in Eq.~(\ref{Eq:Pot_Int}) already appeared in the Lagrangian in Eq.~(\ref{Yuk}). 
The $\lambda_0$ and $\lambda_0'$ terms in the last line in Eq.~(\ref{Eq:Pot_Even}) break the accidental global $U(1)$ symmetries associated with 
the phase transformation of triplet scalar bosons; i.e., $\Delta_{0,1}\to e^{i\theta_{0,1}}\Delta_{0,1} $, so that 
we can avoid the appearance of additional Nambu-Goldstone (NG) bosons. 

The scalar fields can be parameterized as 
\begin{align}
&\Phi =\left[
\begin{array}{c}
\phi^+\\
\phi^0
\end{array}\right],\
\eta =\left[
\begin{array}{c}
\eta^+\\
\eta^0
\end{array}\right],\
\Phi_{3/2} =\left[
\begin{array}{c}
\Phi_{3/2}^{++}\\
\Phi_{3/2}^+
\end{array}\right],\
\Delta_1 = \left[ 
\begin{array}{cc}
\frac{\Delta_1^+}{\sqrt2} & \Delta_1^{++} \\
\Delta_1^0 & -\frac{\Delta_1^+}{\sqrt2}
\end{array}\right],\
\Delta_0 = \left[ 
\begin{array}{cc}
\frac{\Delta_0^0}{\sqrt2} & \Delta_0^+ \\
\bar\Delta_0^- & -\frac{\Delta_0^0}{\sqrt2}
\end{array}\right].   \label{component}
\end{align}
The neutral components of the above fields and the singlet scalar fields can be expressed as
\begin{align}
&S=\frac{1}{\sqrt{2}}(S_R+i S_I),\ 
\chi=\frac{1}{\sqrt{2}}(\chi_R+v_\chi+i \chi_I),\notag\\
&\phi^0=\frac1{\sqrt2}(\phi_R+v_\phi+i\phi_I),\ 
\eta^0=\frac1{\sqrt2}(\eta_R+i\eta_I),\notag\\
&\Delta^0_0=\frac{1}{\sqrt{2}}(\Delta_{0R}+v_{\Delta_0}+i\Delta_{0I}),\
\Delta^0_1=\frac{1}{\sqrt{2}}(\Delta_{1R}+v_{\Delta_1}+i\Delta_{1I}), \label{Eq:neutral}
\end{align}
where $v_\chi,~v_\phi,~v_{\Delta_0}$ and $v_{\Delta_1}$ are the VEVs of 
$\chi$, $\Phi$, $\Delta_0$ and $\Delta_1$, respectively.  
The Fermi constant $G_F$ is given by $v^2\equiv v_\phi^2+2v_{\Delta_1}^2+4v_{\Delta_0}^2=1/(\sqrt{2}G_F)$.
Because $\Delta_0$ is the complex field, $\bar\Delta^+_0$ does not correspond to $(\Delta^-_0)^*$. 

The electroweak rho parameter $\rho$ deviates from unity due to the non-zero value of $v_{\Delta_0}$ and $v_{\Delta_1}$ at the tree level as 
\begin{align}
\rho =\frac{v^2}{v^2+2v_{\Delta_1}^2-4v_{\Delta_0}^2}. \label{Eq:rho}
\end{align}
The experimental value of the rho parameter is close to unity, so that 
the triplet VEVs must be much smaller than $v$ as seen in Eq.~(\ref{Eq:rho}), and
the upper limit is typically given as order of 1 GeV. 

We then calculate the masses of the $\mathbb{Z}_2$-odd scalar bosons which 
are needed to calculate the one-loop diagrams for the lepton masses discussed in the next section.  
In Appendix~A, we also give details of the discussion for mass matrices for the $\mathbb{Z}_2$-even scalar bosons.   

The mass terms for the $\mathbb{Z}_2$-odd scalar bosons can be written by 
\begin{align}
&V_{\text{mass}}=
m^2_{\Phi^{++}_{3/2}}\Phi^{++}_{3/2}\Phi^{--}_{3/2}\notag\\
&+(\Phi^+_{3/2},\eta^+)M_C^2
\left(
\begin{array}{c}
\Phi^-_{3/2}\\
\eta^-
\end{array}
\right)
+\frac{1}{2}(S_I^{},\eta_I^{})M_{I}^2
\left(
\begin{array}{c}
S_I^{}\\
\eta_I^{}
\end{array}
\right)
+\frac{1}{2}(S_R^{},\eta_R^{})M_{R}^2
\left(
\begin{array}{c}
S_R^{}\\
\eta_R^{}
\end{array}
\right), 
\end{align}
where $M_{C}^2$, $M_{I}^2$ and $M_{R}^2$ are the $2\times 2$ mass matrices for the
singly-charged, CP-odd and CP-even scalar boson states, respectively. 
All the masses of $\mathbb{Z}_2$-odd scalar bosons can be extracted from the potential given in Eqs.~(\ref{Eq:Pot_Odd}) and (\ref{Eq:Pot_Int}).  
The mass of the doubly-charged scalar bosons is calculated by
\begin{align}
m^2_{\Phi^{++}_{3/2}}
=
m^2_{3/2}+\frac{1}{2}\left[\kappa_4v^2_\phi+\kappa_{18}v^2_\chi+(\kappa_9-\kappa_{10})v_{\Delta_1}^2+\kappa_{14}v_{\Delta_0}^2\right].
\end{align}
The elements of each mass matrix are obtained as 
\begin{subequations}
\begin{align}
(M_{C}^2)_{11}  &= m_{3/2}^2+\frac{1}{2}\left[(\kappa_3+\kappa_4) v_\phi^2+\kappa_{17}v_\chi^2
+(\kappa_8+\kappa_{9})v_{\Delta_1}^2+\kappa_{13}v_{\Delta_0}^2\right],\\
(M_{C}^2)_{22}  &=\bar{m}^2_\eta-\frac{1}{2}\left(\kappa_2v^2_\phi+\kappa_{16}v^2_\chi+2\kappa_7v_{\Delta_1}^2\right),\\
(M_{C}^2)_{12}  &= -\frac{\kappa_\nu}{\sqrt{2}}v_{\Delta_1}v_{\Delta_0},\\
(M_{I,R}^2)_{11} &=\bar{m}_S^2\mp\kappa_{e1}v^2_\chi,\label{Eq:mass_dif}\\
(M_{I}^2)_{22} &= (M_{R}^2)_{22}=\bar{m}_\eta^2,\\
(M_{I,R}^2)_{12} &= \mp\frac{\kappa_{e2}}{2}v_\phi v_\chi, \label{m12}
\end{align} \label{mass_mat}
\end{subequations}
where 
\begin{align}
\bar{m}_S^2&=m_S^2+\frac{1}{2}\left(\kappa_{5}v^2_\phi+\kappa_{18}v^2_\chi+\kappa_{10}v_{\Delta_1}^2+\kappa_{15}v_{\Delta_0}^2\right),\notag\\
\bar{m}_\eta^2&=m_\eta^2+\frac{1}{2}\left[(\kappa_1+\kappa_{2})v^2_\phi+\kappa_{16}v^2_\chi+(\kappa_{6}+\kappa_{7})v_{\Delta_1}^2+\kappa_{11}v_{\Delta_0}^2\right]. 
\end{align}
The mass eigenstates for the CP-odd and CP-even scalar states are obtained by introducing the mixing angles as
\begin{align}
\left(
\begin{array}{c}
S_I\\
\eta_I
\end{array}\right)=
R(\theta_I)
\left(
\begin{array}{c}
A_1\\
A_2
\end{array}\right),~
\left(
\begin{array}{c}
S_R\\
\eta_R
\end{array}\right)=
R(\theta_R)
\left(
\begin{array}{c}
H_1\\
H_2
\end{array}\right),
~~\text{with}~~R(\theta)=\left(
\begin{array}{cc}
\cos \theta & -\sin \theta \\
\sin \theta & \cos \theta
\end{array}\right). 
\end{align}
The mass eigenvalues and the mixing angles are given as 
\begin{subequations}
\begin{align}
m_{A_{1,2}}^2 &= \frac{1}{2}\left[(M_{I}^2)_{11}+(M_{I}^2)_{22}\pm \sqrt{\left[(M_I^2)_{11}-(M_I^2)_{22}\right]^2+4(M_I^2)_{12}^2}\right],
\\
m_{H_{1,2}}^2 &= \frac{1}{2}\left[(M_{R}^2)_{11}+(M_R^2)_{22}\pm \sqrt{\left[(M_R^2)_{11}-(M_R^2)_{22}\right]^2+4(M_R^2)_{12}^2}\right],\\
\sin2\theta_I&=\frac{2(M_{I}^2)_{12}}{\sqrt{\left[(M_{I}^2)_{11}-(M_{I}^2)_{22}\right]^2+4(M_{I}^2)_{12}^2}}
=\frac{2(M_{I}^2)_{12}}{m_{A_1}^2-m_{A_2}^2}, \\
\sin2\theta_R&=-\frac{2(M_{R}^2)_{12}}{\sqrt{\left[(M_{R}^2)_{11}-(M_{R}^2)_{22}\right]^2+4(M_{R}^2)_{12}^2}}
=\frac{2(M_{R}^2)_{12}}{m_{H_1}^2-m_{H_2}^2}. \label{Eq:m12}
\end{align}
\label{mass_mixing}
\end{subequations}
We note that the mass difference between $H_1$ and $A_1$ and that of $H_2$ and $A_2$ are generated only through the $\kappa_{e1}$ term as seen 
in Eqs.~(\ref{Eq:mass_dif}) and (\ref{Eq:m12}), which is essentially important to obtain 
the non-zero one-loop generated masses of the charged-leptons.  
The mixing angle for the mass matrix $M_C^2$ is also given as 
\begin{align}
\sin2\theta_C&=\frac{2(M_{C}^2)_{12}}{\sqrt{\left[(M_{C}^2)_{11}-(M_{C}^2)_{22}\right]^2+4(M_{C}^2)_{12}^2}}
\simeq -\frac{\sqrt{2}\kappa_\nu v_{\Delta_1}v_{\Delta_0}}{|m_{\Phi_{3/2}^+}^2-m_{\eta^+}^2|}, 
\end{align}
where $m_{\Phi_{3/2}^+}^2=(M_C^2)_{11}$ and $m_{\eta^+}^2=(M_C^2)_{22}$. 
The approximation is valid as long as the triplet VEVs $v_{\Delta_1}$ and $v_{\Delta_0}$ are quite smaller 
than $v_\phi$ and $v_\chi$. 
It is seen that the mixing angle $\theta_C$ is much suppressed by $v_{\Delta_1}$ and $v_{\Delta_0}$, so that 
the mass eigenstates for the singly-charged scalar bosons are almost the same as the corresponding weak eigenstates $\Phi_{3/2}^\pm$ and $\eta^\pm$. 
We note that the lightest neutral scalar boson can be a dark matter candidate. 

In our model, there appears an additional neutral gauge boson, a $Z'$ boson, from the $U(1)'$ gauge symmetry. 
The mass of the $Z'$ boson is given by the VEV of the singlet scalar field $v_\chi$ from the 
kinetic term 
\begin{align}
\mathcal{L}_{\text{kin}}^\chi=
|D_\mu \chi|^2=\Big|[\partial_\mu -ig_{Z'}(x-y)]\chi\Big|^2, 
\end{align}
where $g_{Z'}$ is the $U(1)'$ gauge coupling constant. 
We then obtain the mass of the $Z'$ boson by\footnote{The contribution to $m_{Z'}$ from $v_{\Delta_0}$ and $v_{\Delta_1}$ are neglected. } $m_{Z'}=g_{Z'}|x-y|v_\chi$. 
The $Z'$ mass is constrained by the LEP~II experiment depending on the $U(1)'$ charge of each field~\cite{Carena:2004xs,Gulov:2006gv}.
According to Ref.~\cite{Gulov:2006gv}, the magnitudes of 
the vector coupling $v_\ell$ and the axial vector coupling $a_\ell$ in the $\ell\bar{\ell} Z'$ vertex defined by 
\begin{align}
\mathcal{L}_{\text{int}}=g_{Z'}\bar{\ell}\gamma^\mu(v_\ell-\gamma_5 a_\ell)\ell Z'_\mu, 
\end{align}
are constrained as
\begin{align}
|v_e| < \frac{\sqrt{\pi}m_{Z'}}{g_{Z'}m_Z}\times 0.012,\quad |a_\ell| < \frac{\sqrt{\pi}m_{Z'}}{g_{Z'}m_Z}\times 0.018, \label{const_Zp}
\end{align}
at the 95\% confidence level from the data of $e^+e^-\to e^+e^-$, $e^+e^-\to \mu^+\mu^-$ and $e^+e^-\to \tau^+\tau^-$ processes. 
In our model, $v_\ell$ and $a_\ell$ are given from TABLE~\ref{tab:1} as
\begin{align}
v_\ell = \frac{1}{2}(x+y),\quad a_\ell = \frac{1}{2}(x-y). \label{vell}
\end{align}
The constraint given in Eq.~(\ref{const_Zp}) can be converted into the constraint on $v_\chi$ by using 
Eq.~(\ref{vell}) and the mass formula for $Z'$ as
\begin{align}
v_\chi \gtrsim (2.1~\text{TeV})\times \frac{|x+y|}{|x-y|},\quad v_\chi \gtrsim 1.4~\text{TeV}. 
\end{align}
In the second condition, the dependence of the $U(1)'$ charges is cancelled, so that $v_\chi$ must be larger than 1.4 TeV at least. 
We take $v_\chi=3$ TeV in the numerical analysis discussed in the succeeding sections. 

\section{Observables in the lepton sector}

After the spontaneous electroweak symmetry breaking, 
the mass matrices for the charged-leptons and neutrinos given via the 1-loop diagrams depicted in Fig.~\ref{Fig:Rad_Lep} are obtained by
\begin{align}
&(M_\ell)_{ia}=\sum_{\alpha}
\frac{M_{\alpha}}{64\pi^2}  y_\eta^{i\alpha*}y_S^{a\alpha}
\left[\sin2\theta_R F\left(\frac{m_{H_1}^2}{M_{\alpha}^2},\frac{m_{H_2}^2}{M_{\alpha}^2}\right) 
+
\sin2\theta_I F\left(\frac{m_{A_1}^2}{M_{\alpha}^2},\frac{m_{A_2}^2}{M_{\alpha}^2}\right) \right], \label{m_ell}\\
&(M_\nu)_{ij} =\sum_{\alpha}
\frac{M_{\alpha}}{32\pi^2}  (y_\eta^{i\alpha *} y_{3/2}^{j\alpha}+ y_\eta^{j\alpha *} y_{3/2}^{i\alpha}) 
\sin2\theta_C F\left(\frac{m_{H_{1}^+}^2}{M_{\alpha}^2},\frac{m_{H_2^+}^2}{M_{\alpha}^2}\right), \label{m_nu}
\end{align}
where
\begin{align}
F(x,y)=
\frac{-x\ln x+y\ln y +xy\ln\frac{x}{y}}{(1-x)(1-y)}.
\end{align}
Notice that $(M_\ell)_{ia}$ becomes zero when $m_{H_1}=m_{A_1}$ and $m_{H_2}=m_{A_2}$ are taken, which causes $\sin\theta_R=-\sin\theta_I$ as seen in Eqs.~(\ref{m12}) and (\ref{mass_mixing}).  
Obviously, when $\theta_R=\theta_I=0$ is taken, $(M_\ell)_{ia}$ also is getting to be zero. 
Therefore, both $\kappa_{e1}$ and $\kappa_{e2}$ are required to be non-zero to obtain non-zero masses for the charged-leptons. 
The above mass matrices are diagonalized by introducing the following unitary matrices 
\begin{align}
&U_\ell (M_\ell^\dag M_\ell)U^\dag_\ell={\rm diag}(|m_e|^2,|m_\mu|^2,|m_\tau|^2), \label{Eq:diag_mass} \\
&U_\nu M_\nu U^T_\nu={\rm diag}(m_{\nu_e},m_{\nu_\mu},m_{\nu_\tau}),~~\text{with}~~|U_{\rm PMNS}|\equiv |U^\dag_\ell U_\nu|,\label{def:mns}
\end{align}
where $U_{\rm PMNS}$ is the Pontecorvo-Maki-Nakagawa-Sakata matrix whose elements are given 
from the global fit value of neutrino oscillation data~\cite{Tortola:2012te}. 
In the following, we consider the case with $\alpha=3$. 
In fact, although $\alpha=2$ is enough to obtain two non-zero eigenvalues of $M_\ell$, 
that makes the matrix $U_\ell$ to be not the unit matrix, and 
it causes dangerous LFV processes such as $\mu\to e\gamma$. 
We note that in general, there are $e$-$\tau$ and $\mu$-$\tau$ mixings at the tree level from the $y_\tau^i$ coupling constants and at the one-loop level 
via the $y_\eta^{i\alpha}$ coupling constants.

We here take the following assumptions for the Yukawa coupling constants as 
\begin{align}
&M_1=M_2=M_3=M,\notag\\
&y_\eta=
\begin{pmatrix}
0 & \bar{y}_\eta & \bar{y}_\eta \\ 
\bar{y}_\eta & 0 & \bar{y}_\eta \\
\bar{y}_\eta & \bar{y}_\eta & \bar{y}_\eta
\end{pmatrix}, \quad 
y_S=
\begin{pmatrix}
y_S^{11} & 0 & -y_S^{11} \\ 
0 & y_S^{22} & -y_S^{22} \\
\end{pmatrix},\quad y_\tau^1=y_\tau^2=0 \label{assump}.
\end{align}
Besides, all the elements in $y_\eta$, $y_S$ and $y_{3/2}$ are assumed to be real numbers. 
Under the above assumptions, the mass matrix for the charged-leptons is given as the diagonal form by 
\begin{align}
M_\ell =
\left[
\begin{array}{ccc}
-\tilde{M}_\ell \bar{y}_\eta y_S^{11} & 0 & 0 \\
0 & -\tilde{M}_\ell \bar{y}_\eta y_S^{22} & 0 \\
0 & 0 & \frac{v_\phi}{\sqrt{2}}y_\tau^3 \\
\end{array}\right], \label{Eq:Mell_diag}
\end{align}
where 
\begin{align}
\tilde{M}_\ell\equiv \frac{M}{64\pi^2}
\left[\sin2\theta_R F\left(\frac{m_{H_1}^2}{M^2},\frac{m_{H_2}^2}{M^2} \right)+
\sin2\theta_I F\left(\frac{m_{A_1}^2}{M^2},\frac{m_{A_2}^2}{M^2} \right)\right].  \label{Mtil}
\end{align}
From Eq.~(\ref{Eq:diag_mass}), the (1,1), (2,2) and (3,3) elements in Eq.~(\ref{Eq:Mell_diag}) should correspond to 
$m_e$, $m_\mu$ and $m_\tau$, respectively.  
When we consider the case with $\bar{y}_\eta y_S^{22}$ to be $\mathcal{O}(1)$, $\tilde{M}_\ell$ has to be 
$m_\mu\simeq 0.1$ GeV, which can be achieved by taking $M= \mathcal{O}(1)$ TeV, $\sin2\theta_{R,I} = \mathcal{O}(1)$, and 
$F(x,y)= \mathcal{O}(0.1)$.  
The important point here is that we need almost the maximal mixing between the inert singlet $S$ and doublet $\eta$ fields to reproduce 
the muon mass.
That affects to the DM physics.
Our DM candidate is similar to the property of the isospin inert doublet field due to the maximal mixing. 
One of the allowed region of such a DM mass is known as a resonant solution at around $m_h/2$ if the SM-like Higgs boson is the lightest scalar boson 
among the neutral CP-even neutral bosons, 
in which the DM candidate can satisfy the observed relic density~\cite{Ade:2013lta} and the direct detection~\cite{Aprile:2012nq, Akerib:2013tjd}. 
On the other hand, once the DM mass exceeds the masses of W and Z bosons, 
the annihilation cross section to explain the relic density becomes to be large such as heavier than ${\cal O}$(500) GeV~\cite{Hambye:2009pw}.  
Here we assume that the DM candidate have a mass at around $m_h/2$ so as to increase the testability of the additional charged-leptons $E^\alpha$ which 
can be important to test our model at collider experiments as discussed in the later section.

The mass matrix for neutrinos is expressed by 
\begin{align}
&M_\nu=\tilde{M}_\nu\left[
\begin{array}{ccc}
2(y_{3/2}^{12}+y_{3/2}^{13})& 
(y_{3/2}^{11}+y_{3/2}^{13}+y_{3/2}^{22}+y_{3/2}^{23}) & (\sum_{\alpha}y_{3/2}^{1\alpha}+y_{3/2}^{32}+y_{3/2}^{33}) \\
 & 2(y_{3/2}^{21}+y_{3/2}^{23}) & (\sum_{\alpha}y_{3/2}^{2\alpha}+y_{3/2}^{31}+y_{3/2}^{33})\\
 &  & 2\sum_{\alpha}y_{3/2}^{3\alpha}  \\
\end{array}\right],\\
&\text{with}~~\tilde{M}_\nu\equiv
\bar{y}_\eta\frac{M}{32\pi^2}
\sin2\theta_C F\left(\frac{m_{H_1^+}^2}{M^2},\frac{m_{H_2^+}^2}{M^2} \right). 
\end{align}
The neutrino data given in Eq.~(\ref{def:mns}) can be reproduced by taking appropriate values of $y_{23}^{i\alpha}$ coupling constants. 
The magnitude of the neutrino masses, typically $\mathcal{O}(0.1)$ eV, can be obtained in such a way that 
$\sin2\theta_C$ is taken to be $\mathcal{O}(10^{-11})$ with $\bar{y}_\eta=\mathcal{O}(1)$, $M=\mathcal{O}(1)$ TeV and 
$F=\mathcal{O}(1)$. 
Such a small mixing angle $\theta_C$ can be naturally explained by the smallness of $v_{\Delta_0}$ and $v_{\Delta_1}$, and 
the product of the triplet VEVs should be order of (1 MeV)$^2$ with $\kappa_\nu=\mathcal{O}(1)$. 

\begin{figure}[t]
\begin{center}
 \includegraphics[width=120mm]{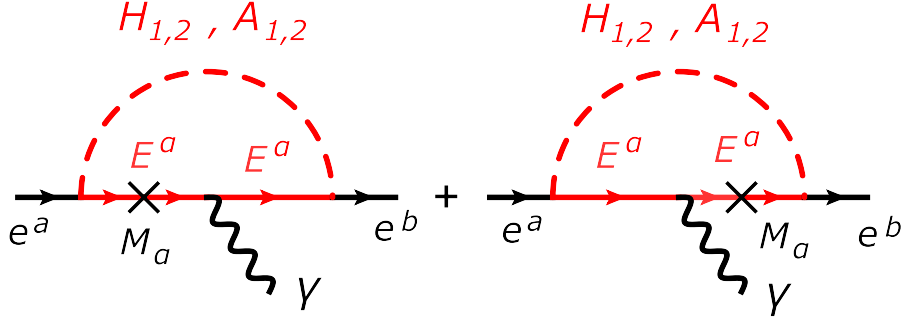}
   \caption{Dominant contributions to the $\ell_a\to \ell_b\gamma$ processes. 
Although the $\Phi_{3/2}^{\pm\pm}$ loop diagram can contribute to the processes, it is neglected, because of the suppression by $m_\mu/M_\alpha$ compared to the dominant contributions.}
   \label{gminus2}
\end{center}
\end{figure}

The muon anomalous magnetic moment has been 
measured at Brookhaven National Laboratory. 
The current average of the experimental results is given by~\cite{bennett}
\begin{align}
a^{\rm exp}_{\mu}=11 659 208.0(6.3)\times 10^{-10},\notag
\end{align}
which has a discrepancy from the SM prediction by $3.2\sigma$~\cite{discrepancy1} to $4.1\sigma$~\cite{discrepancy2} as
\begin{align}
\Delta a_{\mu}=a^{\rm exp}_{\mu}-a^{\rm SM}_{\mu}=(29.0 \pm
9.0\ {\rm to}\ 33.5 \pm
8.2)\times 10^{-10}. \label{g-2_dev}
\end{align}
In our model, the vector-like charged-leptons and $\mathbb{Z}_2$-odd scalar bosons  
can contribute to the $\ell_a\to \ell_b\gamma$ processes 
as shown in Fig.~\ref{gminus2}. 
The amplitude for these processes is calculated by
\begin{align}
\Delta a_{ab}\simeq &\sum_{\alpha=1}^3\frac{m_\mu}{64\pi^2 M_{\alpha}}(y_S^{a\alpha*} y_\eta^{b \alpha}+y_S^{b\alpha*} y_\eta^{a \alpha})
\left[\sin2\theta_RG\left(\frac{m_{H_1}^2}{M_{\alpha}^2},\frac{m_{H_2}^2}{M_{\alpha}^2}\right)
+\sin2\theta_IG\left(\frac{m_{A_1}^2}{M_{\alpha}^2},\frac{m_{A_2}^2}{M_{\alpha}^2}\right)\right], \label{dela}
\\
&\hspace{-1cm}\text{with}~~G(x,y)=\frac{1-4x +3 x^2-2 x^2 \ln x}{2(1-x)^3}-\frac{1-4y +3 y^2-2 y^2 \ln y}{2(1-y)^3}, 
\end{align}
where terms proportional to $(y_S^{a\alpha})^2$ and $(y_{\eta}^{a\alpha})^2$, and 
the $\Phi_{3/2}^{\pm\pm}$ loop contribution
are neglected, because they 
are suppressed by the factor of $m_\mu/M_{\alpha}$ compared to Eq.~(\ref{dela}).
By taking the same assumptions given in Eq.~(\ref{assump}), we obtain
\begin{align}
&\Delta a_{ab}=2\left(\frac{m_\mu}{M^2}\right)\times \mathcal{R} \times (M_\ell)_{ab}, \label{dela2}
\end{align}
where 
\begin{align}
\mathcal{R} \equiv 
\frac{\sin2\theta_RG\left(\frac{m_{H_1}^2}{M^2},\frac{m_{H_2}^2}{M^2}\right)
+\sin2\theta_IG\left(\frac{m_{A_1}^2}{M^2},\frac{m_{A_2}^2}{M^2}\right)}
{\sin2\theta_R F\Big(\frac{m_{H_1}^2}{M^2},\frac{m_{H_2}^2}{M^2} \Big)+
\sin2\theta_I F\Big(\frac{m_{A_1}^2}{M^2},\frac{m_{A_2}^2}{M^2} \Big)}.
\end{align}
The contribution to the muon $g-2$ is given by $\Delta a_{\mu\mu}\equiv \Delta a_\mu$, so that we get the following simple formula
\begin{align}
&\Delta a_\mu=\text{sign}[(M_{\ell})_{\mu\mu}]\times 2\left(\frac{m_\mu}{M}\right)^2\times \mathcal{R}. \label{dela2}
\end{align}
We note that there is no LFV contribution under the assumption in Eq.~(\ref{assump}), because 
the matrix $\Delta a_{ab}$ has a diagonal form. 

In the following, we show numerical calculations for $\tilde{M}_\ell$ and $\Delta a_\mu$ given in Eqs.~(\ref{Mtil}) and (\ref{dela2})
with $\bar{m}_S^2 = \bar{m}_\eta^2~(\equiv \bar{m}^2)$ for simplicity. 
In that case, the masses of neutral $\mathbb{Z}_2$-odd scalar bosons and their mixing angels are given by 
\begin{align}
m_{H_{1,2}}^2 &=\bar{m}^2 +\frac{v_\chi^2}{2}\left(\kappa_{e1}\pm \sqrt{\kappa_{e1}^2 + \kappa_{e2}^2v^2/v_\chi^2 }\right),~
m_{A_{1,2}}^2 =\bar{m}^2 +\frac{v_\chi^2}{2}\left(-\kappa_{e1}\pm \sqrt{\kappa_{e1}^2 + \kappa_{e2}^2v^2/v_\chi^2 }\right),\notag\\ 
\sin2\theta_R &= -\sin2\theta_I = \frac{v}{v_\chi}\frac{\kappa_{e2}}{ \sqrt{\kappa_{e1}^2+\kappa_{e2}^2 v^2/v_\chi^2}}. \label{Z2-mass}
\end{align}
When $\kappa_{e1}$ is taken to be a positive value, 
the mass hierarchy is determined by $m_{H_1}>m_{A_1}>m_{H_2}>m_{A_2}$. 
Therefore, $A_2$ is the lightest neutral $Z_2$-odd particle, and it corresponds to the DM candidate.  
We discuss the case with $\kappa_{e1}>0$ in the following calculations. 
As already explained in the above, 
the mass of the DM candidate should be taken as the half of the Higgs boson mass, so that we take $m_{A_2}=63$ GeV. 
Instead of fixing the physical masses of scalar bosons and mixing angles, we choose $\kappa_{e1}$, $\kappa_{e2}$, $v_\chi$~$(=3$ TeV) , $m_{A_2}~(=63$ GeV) 
as the input parameters. 
In terms of these input variables, we can rewrite the Eq.~(\ref{Z2-mass}) by 
\begin{align}
m_{H_1}^2 &= m_{A_2}^2 + v_\chi^2\left(\kappa_{e1}+ \sqrt{\kappa_{e1}^2 + \kappa_{e2}^2v^2/v_\chi^2 }\right),~
m_{H_2}^2 =  m_{A_2}^2 + v_\chi^2\kappa_{e1},\notag\\
m_{A_1}^2 &=  m_{A_2}^2 + v_\chi^2\sqrt{\kappa_{e1}^2 + \kappa_{e2}^2v^2/v_\chi^2 }.  \label{Z2-mass2}
\end{align}

\begin{figure}[t]
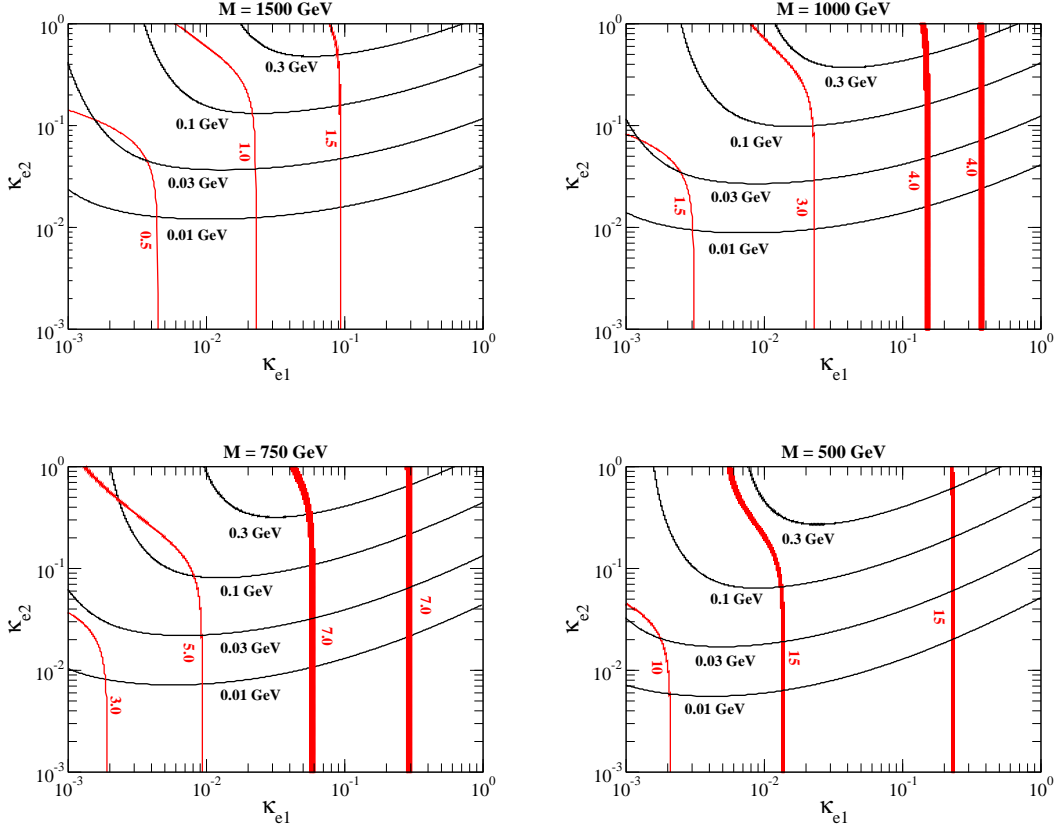

\begin{center}
 \includegraphics[width=65mm]{M_1500.eps}\hspace{8mm}
 \includegraphics[width=65mm]{M_1000.eps}\\ \vspace{8mm}
 \includegraphics[width=65mm]{M_750.eps}\hspace{8mm}
 \includegraphics[width=65mm]{M_500.eps}
   \caption{Contour plots for $\tilde{M}_\ell$ (black curves) and $\Delta a_\mu\times 10^{9}$ (red curves) on the $\kappa_{e1}$-$\kappa_{e2}$ plane.
We take $M=1500$, 1000, 750 and 500 GeV in the upper-left, upper-right, lower-left and lower-right panels, respectively. }
   \label{FIG:k1-k2}
\end{center}
\end{figure}

First in Fig.~\ref{FIG:k1-k2}, 
we show the contours of $\tilde{M}_\ell$ and $\Delta a_\mu\times 10^{9}$ denoted by black and red curves, respectively, on the $\kappa_{e1}$-$\kappa_{e2}$ plane. 
The mass of the vector-like leptons $M$ is taken to be 1500 GeV (upper-left panel), 1000 GeV (upper-right panel) and 750 GeV (lower-left panel) and 
500 GeV (lower-right panel). 
It is seen that the $M$ dependence of $\tilde{M}_\ell$ is weak, while that of $\Delta a_\mu$ is quite strong, 
because of the $(m_\mu/M)^2$ factor in Eq.~(\ref{dela2}).
When we take $\kappa_{e2}\gtrsim 0.01$ ($\kappa_{e2}\gtrsim 0.1$ and $\kappa_{e1}\gtrsim 0.03$), $\tilde{M}_\ell>0.01~(0.1)$ GeV can be obtained. 
Regarding $\Delta a_\mu$, when $M=1500$ GeV and 500 GeV is taken, the value of $\Delta a_\mu$ becomes smaller than about $2.0\times 10^{-9}$
and larger than about $1.0\times 10^{-8}$, respectively, in the region of $\kappa_{e1}$ and $\kappa_{e2}$ shown in Fig.~\ref{FIG:k1-k2}. 
In the case of $M=1000$ (750) GeV, $\Delta a_\mu\simeq 3.0\times 10^{-9}$ is obtained when we take 
$\kappa_{e1}= 0.008$-$0.02$ ($\kappa_{e1}= 0.001$-0.002 and $\kappa_{e2}\lesssim 0.04$). 
Therefore, $M$ to be around 1 TeV is favored by the measurement of the muon $g-2$. 

\begin{figure}[t]
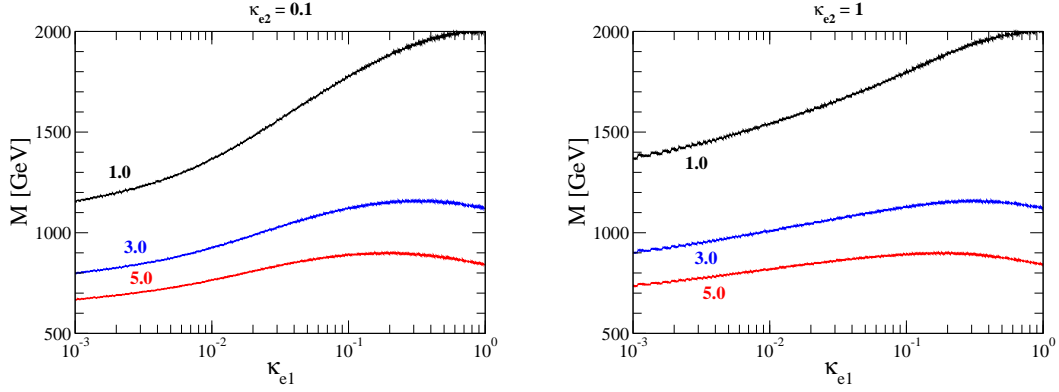

\begin{center}
 \includegraphics[width=65mm]{g2_k2_01.eps}\hspace{8mm}
 \includegraphics[width=65mm]{g2_k2_1.eps}\\ \vspace{8mm}
   \caption{Contour plots for $\Delta a_\mu\times 10^{9}$ on the $\kappa_{e1}$-$M$ plane.  
The value of $\kappa_{e2}$ is fixed to be 0.1 and 1 in the left and right panels, respectively.}
   \label{FIG:k1-M}
\end{center}
\end{figure}

Second, in Fig.~\ref{FIG:k1-M}, we show the contours of $\Delta a_\mu\times 10^{9}$ on the $\kappa_{e1}$-$M$ plane for the fixed 
values of $\kappa_{e2}$ to be 0.1 (left panel) and 1 (right panel). 
We can expect that the result does not change in the case with values of $\kappa_{e2}$ smaller than 0.1 as seen in Fig.~\ref{FIG:k1-k2}. 
In both cases with $\kappa_{e2}=0.1$ and 1, when $M$ is taken to be around 1000 GeV, we can get $\Delta a_\mu\simeq 3.0\times 10^{-9}$. 
When $M$ is taken to be smaller (larger) than about 700 (2000) GeV, 
$\Delta a_\mu$ becomes larger (smaller) than 5.0~(1.0)$\times 10^{-9}$, which are outside of the 2-sigma error of the measured $\Delta a_\mu$. 

\begin{figure}[!t]
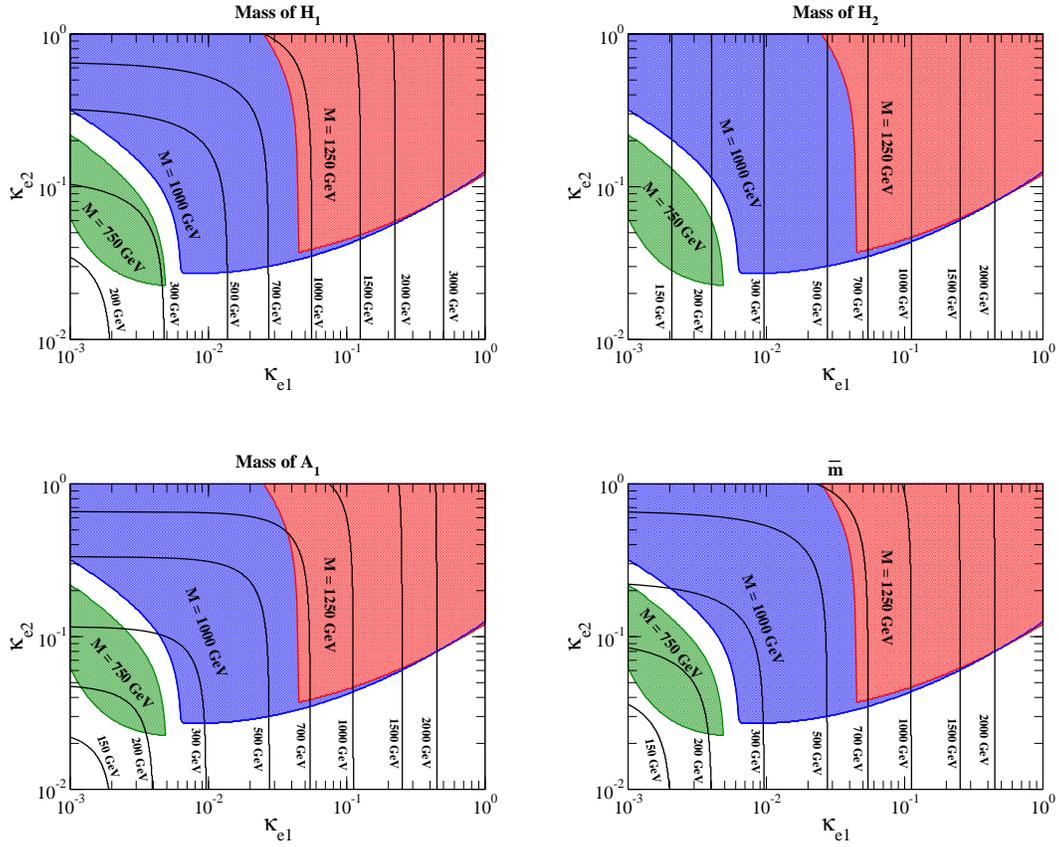

\begin{center}
 \includegraphics[width=65mm]{mH1.eps}\hspace{8mm}
 \includegraphics[width=65mm]{mH2.eps}\\ \vspace{8mm}
 \includegraphics[width=65mm]{mA1.eps}\hspace{8mm}
\includegraphics[width=65mm]{mbar.eps}
   \caption{Contour plots for $m_{H_1}$ (upper-left panel), $m_{H_2}$ (upper-right panel), 
$m_{A_1}$ (lower-left panel) and $\bar{m}$ (lower-right panel) on the $\kappa_{e1}$-$\kappa_{e2}$ plane. 
The red and blue shaded regions satisfy $\tilde{M}_\ell>0.03$ GeV, $2.0\times 10^{-9}<\Delta a_\mu<4.2\times 10^{-9}$. }
   \label{FIG:Mass}
\end{center}
\end{figure}

Finally, in Fig.~\ref{FIG:Mass}, we show the contour plots for the output values of 
$m_{H_1}$ (upper-left), $m_{H_2}$ (upper-right), $m_{A_1}$ (lower-left) and 
$\bar{m}$ (lower-right) on the $\kappa_{e1}$-$\kappa_{e2}$ plane.  
The red, blue and green shaded regions satisfy $\tilde{M}_\ell>0.03$ GeV and $2.0\times 10^{-9}<\Delta a_\mu<4.2\times 10^{-9}$ in the case of 
$M=1250$, 1000 and 750 GeV, respectively. 
The condition for $\Delta a_\mu$ corresponds to the requirement where the prediction of $\Delta a_\mu$ is inside the 1-sigma 
error from the measurement. 
We can see that the shaded regions are shifted from the lower-left region to the upper-right region on the $\kappa_{e1}$-$\kappa_{e2}$ plane when 
the value of $M$ is changed from 750 GeV to 1250 GeV.
That can be understood in such a way that the prediction of $\Delta a_\mu$ is getting smaller as $M$ is increased, so that 
to compensate the suppression by $M$ we need larger values of $\kappa_{e1}$ and $\kappa_{e2}$.  
In the case of $M=750$ GeV,
the case with $220\lesssim m_{H_1} \lesssim 400$ GeV, 
$180\lesssim m_{A_1} \lesssim 400$ GeV and $110\lesssim m_{H_2} \lesssim 220$ GeV is favored by $g-2$ data and 
natural explanation for the muon mass by meaning of $\tilde{M}_\ell>0.03$ GeV. 
Similarly in the case of $M=1000$ GeV, 
the case of $ m_{H_1} \gtrsim 350$ GeV, $ m_{A_1} \gtrsim 250$ GeV and $ m_{A_1} \gtrsim 250$ GeV is favored. 
\section{ Collider phenomenology}

\begin{center}
\begin{table}[t]
\begin{tabular}{c||c|c|c|c||c|c|c|c|c|c}\hline\hline 
 & \multicolumn{4}{c||}{Input parameters}& \multicolumn{6}{c}{Outputs}  \\\hline
 & $M$ & $m_{A_2}$ & $\kappa_{e1}$  & $\kappa_{e2}$ & $m_{H_1}$ & $m_{A_1}$ & $m_{H_2}$& $\bar{m}$ & $\tilde{M}_\ell$ & $\Delta a_\mu$  \\\hline
Benchmark I & 750 GeV & 63 GeV & 0.001 & $-0.1$ & 295 GeV & 279 GeV & 113 GeV &213 GeV& $0.036$ GeV & 3.6$\times 10^{-9}$   \\\hline
Benchmark II  & 1000 GeV & 63 GeV & 0.03 & $-0.1$ & 744 GeV & 533 GeV & 523 GeV &528 GeV& $0.097$ GeV & 3.2$\times 10^{-9}$   \\\hline\hline
\end{tabular}
\caption{Benchmark input parameters and corresponding outputs.   }
\label{tab:bm}
\end{table}
\end{center}

In this section, we discuss the collider phenomenology in our model.  
As seen in Feynman diagrams shown in Fig.~\ref{Fig:Rad_Lep}, the vector-like leptons $E_\alpha$~($\alpha=1,2,3$) 
play an essential role to generate the masses of both charged-leptons and neutrinos. 
Therefore, the detection of $E_\alpha$ is important to test our model.  
In the following, we discuss the decay property of $E_\alpha$ based on the assumptions given in Eq.~(\ref{assump}) and the mass formulae 
expressed in Eq.~(\ref{Z2-mass2}). 

The decay of $E_\alpha$ depends on the mass spectrum for the $\mathbb{Z}_2$-odd particles, some of which can be determined by taking into 
account the muon mass and $\Delta a_\mu$ as seen in Fig.~\ref{FIG:Mass}. 
As examples, we consider the benchmark points as listed in Table~\ref{tab:bm}.  
Among the $\mathbb{Z}_2$-odd particles, the mass of $\Phi_{3/2}^{\pm\pm}$ as well as $\Phi_{3/2}^{\pm}$ are not so constrained from experiments. 
We can then consider the two cases; (1) $M<m_{\Phi_{3/2}^{++}}$ and (2) $M>m_{\Phi_{3/2}^{++}}$. 
In the case (1), $E_\alpha^\pm$ can decay into $\ell^\pm \varphi^0$ and $\eta^\pm \nu$, where 
$\varphi^0$ denotes the neutral $\mathbb{Z}_2$-odd scalar boson ($H_1$, $H_2$, $A_1$ and $A_2$). 
On the other hand, in the case (2), $E_\alpha^\pm$ can decay into
$\Phi_{3/2}^{\pm\pm}\ell^\mp$ and $\Phi_{3/2}^{\pm}\nu$ via the $y_{3/2}$ couplings in 
addition to the decay modes discussed in the case (1). 

If the case (1) is realized, the decay of $E_\alpha$ is getting more sensitive to the structure of Yukawa interactions related 
to the generation of the charged-lepton masses, and we concentrate on this case. 
Because of the assumption $M_1=M_2=M_3=M$, the Drell-Yan pair production cross sections for $E_1$, $E_2$ and $E_3$ 
are the same, 
so that the sum of the decay rates of $E_\alpha$ defined by $\Gamma(E\to X)\equiv \sum_\alpha \Gamma(E_\alpha \to X)$ can be measured. 
They are calculated as 
\begin{align}
\Gamma(E^\pm \to e^\pm \varphi_i) &=\frac{M}{32\pi}\bar{y}_\eta^2c_{\eta_i}^2\left(1-\frac{m_{\varphi_i}^2}{M^2}\right)^2,\\
\Gamma(E^\pm \to \tau^\pm \varphi_i)&=\frac{3M}{64\pi}\bar{y}_\eta^2c_{\eta_i}^2\left(1-\frac{m_{\varphi_i}^2}{M^2}\right)^2,\\
\Gamma(E^\pm \to \mu^\pm \varphi_i) &=
\frac{M}{32\pi}[\bar{y}_\eta^2c_{\eta_i}^2+(y_S^{22})^2c_{S_i}^2]\left(1-\frac{m_{\varphi_i}^2}{M^2}\right)^2,
\label{Edecay_mu}\\
\Gamma(E^\pm \to \eta^\pm\nu ) &=
\frac{7M}{64\pi}\bar{y}_\eta^2\left(1-\frac{m_{\eta^+}^2}{M^2}\right)^2,
 \end{align}
where 
\begin{align}
&\varphi_i= (H_1,H_2,A_1,A_2),~~
c_{\eta_i}= (\sin\theta_R,\cos\theta_R,\sin\theta_I,\cos\theta_I),\notag\\
&c_{S_i}= (\cos\theta_R,\sin\theta_R,\cos\theta_I,\sin\theta_I),
~\text{for}~(i=1,2,3,4). 
\end{align}
For the $E^\pm \to \eta^\pm \nu $ mode, final states with $\nu_e$, $\nu_\mu$ and $\nu_\tau$ are summed. 
We note that the decay branching fraction of $E^\pm$ can be described by the ratio of the two Yukawa couplings 
$r\equiv y_S^{22}/\bar{y}_\eta$ instead of $y_S^{22}$ and $\bar{y}_\eta$. 
From Eq.~(\ref{Eq:Mell_diag}), the product $y_S^{22}\times\bar{y}_\eta$ has to be fixed to obtain the correct muon mass, i.e., 
$y_S^{22}\times\bar{y}_\eta\simeq 1.0~(2.7)$ in Benchmark~I (II). 
Therefore, each of Yukawa couplings is determined by 
$y_S^{22}=\sqrt{2.94r}$ ($\sqrt{1.09r}$) and $\bar{y}_\eta=1/\sqrt{r}$ ($\sqrt{2.75/r}$) in Benchmark~I (Benchmark~II). 

\begin{figure}[t]
\begin{center}
 \includegraphics[width=100mm]{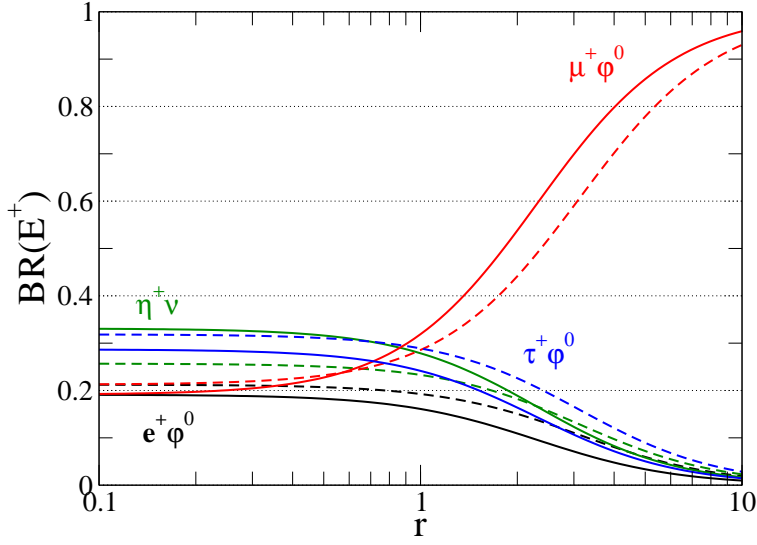}
   \caption{Branching fractions of $E^+$ as a function of the ratio $r\equiv y_S^{22}/\bar{y}_\eta$. 
The mass of $\eta$ is taken to be $\bar{m}$. 
The solid and dashed curves are respectively the results in the Benchmark~I and Benchmark~II listed in TABLE~\ref{tab:bm}. 
$\ell^+\varphi^0$ denotes the sum of all the possible modes with a neutral $\mathbb{Z}_2$-odd scalar boson $\varphi^0$ and a charged-lepton 
$\ell$ ($=e,\mu,\tau$). }
   \label{BR1}
\end{center}
\end{figure}

In Fig.~\ref{BR1}, we show the decay branching fraction of $E^+$ as a function of the ratio $r$.
The decay processes with a neutral $\mathbb{Z}_2$-odd scalar boson in the final state are summed. 
The results in Benchmark~I and Benchmark~II are shown as the solid and dashed curves, respectively. 
We can see that in the large $r$ region, only the $\mu^+\varphi^0$ mode is increased compared to all the other channels, 
because only the $E^\pm\mu^\mp \varphi^0$ vertex is enhanced by the large $y_{S}^{22}$. 
We note that $H_1$ and $H_2$ ($A_1$) can further decay into $A_2Z^{(*)}$ $(H_2Z^{(*)})$, and $\eta^\pm$ can decay into 
$W^\pm A_2$. 
Therefore, except for the $E^\pm \to \ell^\pm A_2$ channel, $E^\pm$ decay into a charged-lepton plus $A_2$ associated with 
weak gauge bosons. 
 
Finally, we would like to comment on the phenomenology of two pairs of doubly-charged scalar bosons; i.e., 
$\Delta_1^{\pm\pm}$ and $\Phi_{3/2}^{\pm\pm}$, which is also important in addition to that of $E_\alpha^\pm$ in order to identify our model. 
In our model, the tree level Yukawa interaction for $\Delta_1^{\pm\pm}$ is forbidden by the $U(1)'$ symmetry, 
so that $\Delta_1^{\pm\pm}$ cannot decay into the same sign dilepton unlike the minimal Higgs triplet model (HTM)~\cite{typeII}. 
Thus, the main decay mode of $\Delta_1^{\pm\pm}$ is the same sign diboson; i.e., $\Delta_1^{\pm\pm}\to W^\pm W^\pm$ as long as 
$\Delta_1^{\pm\pm}$ are the lightest of all the other component fields in $\Delta_1$. 
Such a decay mode can appear in  the HTM with the case of a rather large triplet VEV; i.e., about larger than 1 MeV. 
The collider phenomenology for the diboson decay scenario had been discussed in Refs.~\cite{diboson,diboson2}. 
In Ref.~\cite{diboson2}, by using the LHC data with the collision energy of 7 TeV and integrated luminosity of 4.7 fb$^{-1}$,  
the lower mass bound for the doubly-charged scalar bosons has been found to be about 60 GeV.
On the other hand, the following decay modes of $\Phi_{3/2}^{\pm\pm}$ via the $\kappa_\nu$ or $\tilde{\kappa}_\nu$ couplings
are possible\footnote{These decay modes can be replaced by $\Phi_{3/2}^{\pm\pm}\to W^{\pm *}\Phi_{3/2}^{\pm}$ if 
there is a mass difference between $\Phi_{3/2}^{\pm\pm}$ and $\Phi_{3/2}^{\pm}$.  } as long as they are kinematically allowed; 
$\eta^\pm \Delta_0^\pm$, $\eta^\pm \bar{\Delta}_0^{\pm}$, $\eta^\pm \Delta_1^\pm$ and $\eta^0 \Delta_1^{\pm\pm}$.  
If we consider the case (1) mentioned in the above, $\Phi_{3/2}^{\pm\pm}$ can mainly decay into a charged-lepton plus $E_\alpha^\pm$.  
In that case, all the decay branching fractions for the modes with the component of the triplet Higgs field $\Delta_1$ or $\Delta_0$ are quite suppressed 
due to the small triplet VEV.

\section{Conclusions and Discussions }

We have modified our previous model in Ref.~\cite{Okada:2013iba} with the one-loop generation of masses for neutrinos, muon and electron. 
The masses of muon and electron are induced by the $\bar{L}_Le_R\Phi\chi$ operator, and those of neutrinos are 
generated from the $\overline{L_L^c}L_L\Delta_0\Delta_1$ operator. 
The doublet VEV $v_\phi$ and singlet VEV $v_\chi$ are respectively determined by the Fermi constant $G_F$ and the $Z'$ mass which 
have to be above 1 TeV from the constraint from LEP~II experiment. 
On the other hand, the triplet VEVs must be smaller than about 1 GeV due to the constraint from the electroweak rho parameter. 
Therefore, the mass hierarchy between neutrinos and charged-leptons can be naturally described by the suppression of the triplet VEVs for the neutrino masses 
with ${\cal O}$(1) Yukawa coupling constants $y_S^{22}$ and $\bar{y}_\eta$. 

In our model, the lightest $\mathbb{Z}_2$-odd neutral scalar boson can be a DM candidate, and $A_2$ corresponds to it when $\kappa_{e1}$ is taken to be 
a positive value. 
The mass of DM is set to be $m_h/2$ in order to satisfy the relic density and the constraint from the direct detection. 

Under the assumptions given in Eq.~(\ref{assump}) and the requirement from the DM physics, 
we have calculated the muon $g-2$ and $\tilde{M}_\ell$. 
The results of favored parameter regions by taking into account the above observables can be seen in Fig.~\ref{FIG:Mass}. 
We have found that the mass of vector-like charged-leptons $E_\alpha^\pm$ to be around 1 TeV is favored by taking into account the above 
observables. 

We have studied the decay property of $E_\alpha^\pm$ whose decay branching fractions are shown in Fig.~\ref{BR1} with the two benchmark 
parameter sets in the case of $M<m_{\Phi_{3/2}^{++}}$. 
We have found that $E_\alpha^\pm$ can mainly decay into a muon plus neutral $\mathbb{Z}_2$-odd scalar bosons which can further decay into 
the Z boson and the DM candidate $A_2$ when $y_{S}^{22}$ is relatively larger than $\bar{y}_\eta$. 
\\\\
\noindent
{$Acknowledgments$}:
\\
The authors would like to thank Hiroaki Sugiyama for an useful comment on the model building. 
K.Y. was supported in part by the National Science Council of R.O.C. under Grant No. NSC-101-2811-M-008-014.

\begin{appendix}

\section{Higgs sector with the $\mathbb{Z}_2$ even part}

In this Appendix, we give the mass formulae for the $\mathbb{Z}_2$-even Higgs bosons, in which we neglect the terms proportional to $v_{\Delta_1}^2$, $v_{\Delta_0}^2$, and $v_{\Delta_0}v_{\Delta_1}$. 

From the tadpole conditions, the scalar invariant mass parameters ${m}_\Phi^2$, $m_\chi^2$, $m_{\Delta_1}^2$ and $m_{\Delta_0}^2$
can be rewritten by 
\begin{align}
m_\Phi^2& \simeq  \lambda_0v_{\Delta_1}v_\chi+\frac{\lambda'_0}{\sqrt{2}}v_{\Delta_0}v_\chi-\lambda_1v_\phi^2-\frac{\lambda_{12}}{2}v_\chi^2,\notag\\
m_\chi^2&\simeq\frac{1}{2}\left(\lambda_0\frac{v_{\Delta_1}}{v_\chi}+\frac{\lambda'_0}{\sqrt{2}}\frac{v_{\Delta_0}}{v_\chi}-\lambda_{11}\right)v_\phi^2-\lambda_{6}v_\chi^2,
\notag\\
m_{\Delta_0}^2& \simeq
\frac{1}{2}\left(\frac{\lambda'_0v_\chi}{\sqrt{2}v_{\Delta_0}}-\lambda_9\right)v_\phi^2-\frac{\lambda_{13}}{2}v_\chi^2,\notag\\
m_{\Delta_1}^2 &\simeq 
\frac{1}{2}\left(\lambda_0 \frac{v_\chi}{v_{\Delta_1}}- \bar\lambda\right)v_\phi^2-\frac{\lambda_{12}}{2}v_\chi^2,
\end{align}
where $\bar\lambda\equiv\lambda_7+\lambda_8$.

There are one pair of doubly-charged states, four pairs of singly-charged states, four CP-odd and four CP-even states. 
First, the mass of the doubly-charged Higgs boson $\Delta^{\pm\pm}$ is calculated as 
\begin{align}
m_{\Delta_1^{++}}^2\simeq \left(\frac{\lambda_0}{2}\frac{v_\chi}{v_{\Delta_1}}-\lambda_8\right)v_\phi^2.
\end{align}

Second, we discuss the masses of the singly-charged scalar states. 
Because one of four pairs corresponds to the NG boson state which are absorbed into the longitudinal component of the $W$ boson, 
we can obtain the block diagonal form of the matrix, in which the NG mode is separated into the physical singly-charged scalar bosons as. 
When we define the mass matrix for the singly-charged state in the basis of $(\phi^+,\Delta_1^+,\Delta_0^+,\bar{\Delta}_0^{+})$ 
with $\bar{\Delta}_0^{+}=\bar{\Delta}_0^{-*}$ as
\begin{align}
O_C^T \bar{M}_{C}^2 O_C\simeq  \left[
\begin{array}{cccc}
0&0&0&0\\
& \frac{\sqrt{2}\lambda_0'}{4}\frac{v_\phi^2 v_\chi}{v_{\Delta_0}}& 0 & \frac{\lambda_{10}}{2}v_\phi^2\\
&&\frac{1}{2}\left(\frac{v_\chi}{v_{\Delta_1}}\lambda_0-\lambda_8\right)v_\phi^2 & -\lambda_0' v_{\Delta_1}v_\chi \\
&&& \frac{\lambda_0'}{2\sqrt{2}}\frac{v_\phi^2 v_\chi}{v_{\Delta_0}}
\end{array}\right], 
\end{align}
where 
\begin{align}
O_C\simeq  \left[
\begin{array}{cccc}
-1 & 0 & \frac{\sqrt{2}v_{\Delta_1}}{v_\phi} & -\frac{2v_{\Delta_1}}{v_\phi}\\
-\frac{\sqrt{2}v_{\Delta_1}}{v_\phi} & 0 & -1 & 0\\ 
\frac{\sqrt{2}v_{\Delta_0}}{v_\phi} & \frac{1}{\sqrt{2}} & 0 & -\frac{1}{\sqrt{2}}\\
\frac{\sqrt{2}v_{\Delta_0}}{v_\phi} & -\frac{1}{\sqrt{2}} & 0 & -\frac{1}{\sqrt{2}}\\
\end{array}\right], 
\end{align}

Third, two of the four CP-odd states correspond to the NG bosons which are absorbed by the longitudinal component of 
the $Z$ boson and additional neutral gauge boson from the $U(1)'$ symmetry. 
Therefore, the mass matrix for the CP-odd states in the basis of $(\phi_I,\chi_I,\Delta_{1I},\Delta_{0I})$ as can be expressed by 
the block diagonal form with $2\times 2$ submatrix to be non-zero as
\begin{align}
O_{I}^T \bar{M}_I^2 O_{I} \simeq \left[
\begin{array}{cccc}
0&0&0&0\\
0&0&0&0\\
&&\frac{\lambda_0'}{2\sqrt{2}}\frac{v_\phi^2v_\chi}{v_{\Delta_0}} & \lambda_0 v_\phi v_{\Delta0} \\
&&& \frac{\lambda_0}{2}\frac{v_\phi^2v_\chi}{v_{\Delta_1}}
\end{array}\right],
\end{align}
where 
\begin{align}
O_{I} \simeq  \left[
\begin{array}{cccc}
\frac{1}{\sqrt{2}}r_+ & -\frac{1}{\sqrt{2}}r_-  &0&\frac{2v_{\Delta_1}}{v_\phi}\\
-\frac{\sqrt{2}}{r_+}\frac{v_\chi}{\bar{v}}&-\frac{\sqrt{2}}{r_-}\frac{v_\chi}{\bar{v}}&\frac{v_{\Delta_0}}{v_\chi}&\frac{v_{\Delta_1}}{v_\chi}\\
\frac{\sqrt{2}}{r_+}\frac{v_{\Delta_1}}{v_\phi}&-\frac{\sqrt{2}}{r_-}\frac{v_{\Delta_1}}{v_\phi}&0&-1\\
\frac{\sqrt{2}}{r_+}\frac{v_{\Delta_0}}{\bar{v}}&-\frac{\sqrt{2}}{r_-}\frac{v_{\Delta_0}}{\bar{v}}&1&0
\end{array}\right],
\end{align}
with $r_\pm\equiv \sqrt{1\pm 1/\sqrt{1+4v_\chi^2/v_\phi^2}}$. 

Finally, the mass matrix for the CP-even states in the basis of $(\phi_R,\chi_R,\Delta_{1R},\Delta_{0R})$ is given by  
\begin{align}
&  \bar{M}_R^2 \simeq \nn\\
&\left[
\begin{array}{cccc}
2\lambda_1  v_\phi^2
& \left(\lambda_{11}v_\chi-\lambda_0v_{\Delta_1}-\frac{\lambda_0'}{\sqrt{2}}v_{\Delta_0}\right)v_\phi
& \left[-\lambda_0v_\chi+(\lambda_7+\lambda_8 )v_{\Delta_1}\right]v_\phi
& \left(-\frac{\lambda_0'}{\sqrt{2}}v_\chi+\lambda_9v_{\Delta_0}\right)v_\phi \\
 &\frac{1}{2}\left(4\lambda_6 v_\chi^2+\lambda_0\frac{v_{\Delta_1}v_\phi^2}{v_\chi}+\frac{\lambda_0'}{\sqrt{2}}\frac{v_{\Delta_0} v_\phi^2}{v_\chi}\right)
& \lambda_{12}v_\chi v_{\Delta_1}-\frac{\lambda_0}{2}v_\phi^2 & \lambda_{13}v_\chi v_{\Delta_0}-\frac{\lambda_0'}{2\sqrt{2}}v_\phi^2 \\
&& \frac{\lambda_0}{2}\frac{v_\chi v_\phi^2}{v_{\Delta_1}}& 0\\
&&&\frac{\sqrt{2}\lambda_0'}{4}\frac{v_\phi^2v_\chi}{v_{\Delta_0}}
\end{array}\right].
\end{align}

\end{appendix}


\end{document}